



\message{ Assuming 8.5" x 11" paper }    

\magnification=\magstep1	          
\vsize=8.7 true in
\raggedbottom

\parskip=9pt

\def\singlespace{\baselineskip=12pt}      
\def\sesquispace{\baselineskip=16pt}      




\def\author#1 {\medskip\centerline{\it #1}\bigskip}
\font\titlefont=cmb10 scaled\magstep2 
\def\address#1{\centerline{\it #1}\smallskip}
\def\email#1{\smallskip\centerline{\it address for email: #1}} 

\def\AbstractBegins
{
 \singlespace                                        
 \bigskip\leftskip=1.5truecm\rightskip=1.5truecm     
 \centerline{\bf Abstract}
 \smallskip
 \noindent	
 } 
\def\AbstractEnds
{
 \bigskip\leftskip=0truecm\rightskip=0truecm       
 }

\def\section #1 {\bigskip\noindent{\headingfont #1 }\par\nobreak\smallskip\noindent}

\def\linebreak{\hfil\break}
\def\lbr{\linebreak}

\def\ReferencesBegin
{
 \singlespace					   
 \vskip 0.5truein
 \centerline           {\bf References}
 \par\nobreak
 \medskip
 \noindent
 \parindent=2pt
 \parskip=6pt			
 }

\def\reference{\hangindent=1pc\hangafter=1} 

\def\ref{\reference}
\def\journaldata#1#2#3#4{{\it #1\/}\phantom{--}{\bf #2$\,$:} $\!$#3 (#4)}
\def\eprint#1{{\tt #1}}
\def\arxiv#1{\hbox{\tt http://arXiv.org/abs/#1}}

\font\titlefont=cmb10 scaled\magstep2 
\font\headingfont=cmb10 at 12pt

\def\sepref{\parskip=4pt \par \hangindent=1pc\hangafter=0}
 %





\edef\resetatcatcode{\catcode`\noexpand\@\the\catcode`\@\relax}
\ifx\miniltx\undefined\else\endinput\fi
\let\miniltx\box

\def\makeatletter{\catcode`\@11\relax}

\makeatletter

\def\@makeother#1{\catcode`#1=12\relax}

\def\@ifnextchar#1#2#3{%
  \let\reserved@d=#1%
  \def\reserved@a{#2}\def\reserved@b{#3}%
  \futurelet\@let@token\@ifnch}
\def\@ifnch{%
  \ifx\@let@token\@sptoken
    \let\reserved@c\@xifnch
  \else
    \ifx\@let@token\reserved@d
      \let\reserved@c\reserved@a
    \else
      \let\reserved@c\reserved@b
    \fi
  \fi
  \reserved@c}
\begingroup
\def\:{\global\let\@sptoken= } \:  
\def\:{\@xifnch} \expandafter\gdef\: {\futurelet\@let@token\@ifnch}
\endgroup

\def\@ifstar#1{\@ifnextchar *{\@firstoftwo{#1}}}
\long\def\@dblarg#1{\@ifnextchar[{#1}{\@xdblarg{#1}}}
\long\def\@xdblarg#1#2{#1[{#2}]{#2}}

\long\def \@gobble #1{}
\long\def \@gobbletwo #1#2{}
\long\def \@gobblefour #1#2#3#4{}
\long\def\@firstofone#1{#1}
\long\def\@firstoftwo#1#2{#1}
\long\def\@secondoftwo#1#2{#2}

\def\NeedsTeXFormat#1{\@ifnextchar[\@needsf@rmat\relax}
\def\@needsf@rmat[#1]{}
\def\ProvidesPackage#1{\@ifnextchar[%
    {\@pr@videpackage{#1}}{\@pr@videpackage#1[]}}
\def\@pr@videpackage#1[#2]{\wlog{#1: #2}}

\let\DeclareOption\@gobbletwo
\def\ProcessOptions{\@ifstar\relax\relax}

\def\RequirePackage{%
  \@fileswithoptions\@pkgextension}
\def\@fileswithoptions#1{%
  \@ifnextchar[
    {\@fileswith@ptions#1}%
    {\@fileswith@ptions#1[]}}
\def\@fileswith@ptions#1[#2]#3{%
  \@ifnextchar[
  {\@fileswith@pti@ns#1[#2]#3}%
  {\@fileswith@pti@ns#1[#2]#3[]}}

\def\@fileswith@pti@ns#1[#2]#3[#4]{%
    \def\reserved@b##1,{%
      \ifx\@nil##1\relax\else
        \ifx\relax##1\relax\else
         \noexpand\@onefilewithoptions##1[#2][#4]\noexpand\@pkgextension
        \fi
        \expandafter\reserved@b
      \fi}%
      \edef\reserved@a{\zap@space#3 \@empty}%
      \edef\reserved@a{\expandafter\reserved@b\reserved@a,\@nil,}%
  \reserved@a}

\def\zap@space#1 #2{%
  #1%
  \ifx#2\@empty\else\expandafter\zap@space\fi
  #2}

\let\@empty\empty
\def\@pkgextension{sty}

\def\@onefilewithoptions#1[#2][#3]#4{%
  \input #1.#4 }

\def\typein{%
  \let\@typein\relax
  \@testopt\@xtypein\@typein}
\def\@xtypein[#1]#2{%
  \message{#2}%
  \advance\endlinechar\@M
  \read\@inputcheck to#1%
  \advance\endlinechar-\@M
  \@typein}
\def\@namedef#1{\expandafter\def\csname #1\endcsname}
\def\@nameuse#1{\csname #1\endcsname}
\def\@cons#1#2{\begingroup\let\@elt\relax\xdef#1{#1\@elt #2}\endgroup}
\def\@car#1#2\@nil{#1}
\def\@cdr#1#2\@nil{#2}
\def\@carcube#1#2#3#4\@nil{#1#2#3}
\def\@preamblecmds{}

\def\@star@or@long#1{%
  \@ifstar
   {\let\l@ngrel@x\relax#1}%
   {\let\l@ngrel@x\long#1}}

\let\l@ngrel@x\relax
\def\newcommand{\@star@or@long\new@command}
\def\new@command#1{%
  \@testopt{\@newcommand#1}0}
\def\@newcommand#1[#2]{%
  \@ifnextchar [{\@xargdef#1[#2]}%
                {\@argdef#1[#2]}}
\long\def\@argdef#1[#2]#3{%
   \@ifdefinable #1{\@yargdef#1\@ne{#2}{#3}}}
\long\def\@xargdef#1[#2][#3]#4{%
  \@ifdefinable#1{%
     \expandafter\def\expandafter#1\expandafter{%
          \expandafter
          \@protected@testopt
          \expandafter
          #1%
          \csname\string#1\expandafter\endcsname
          {#3}}%
       \expandafter\@yargdef
          \csname\string#1\endcsname
           \tw@
           {#2}%
           {#4}}}
\def\@testopt#1#2{%
  \@ifnextchar[{#1}{#1[#2]}}
\def\@protected@testopt#1{
  \ifx\protect\@typeset@protect
    \expandafter\@testopt
  \else
    \@x@protect#1%
  \fi}
\long\def\@yargdef#1#2#3{%
  \@tempcnta#3\relax
  \advance \@tempcnta \@ne
  \let\@hash@\relax
  \edef\reserved@a{\ifx#2\tw@ [\@hash@1]\fi}%
  \@tempcntb #2%
  \@whilenum\@tempcntb <\@tempcnta
     \do{%
         \edef\reserved@a{\reserved@a\@hash@\the\@tempcntb}%
         \advance\@tempcntb \@ne}%
  \let\@hash@##%
  \l@ngrel@x\expandafter\def\expandafter#1\reserved@a}
\long\def\@reargdef#1[#2]#3{%
  \@yargdef#1\@ne{#2}{#3}}
\def\renewcommand{\@star@or@long\renew@command}
\def\renew@command#1{%
  {\escapechar\m@ne\xdef\@gtempa{{\string#1}}}%
  \expandafter\@ifundefined\@gtempa
     {\@latex@error{\string#1 undefined}\@ehc}%
     {}%
  \let\@ifdefinable\@rc@ifdefinable
  \new@command#1}
\long\def\@ifdefinable #1#2{%
      \edef\reserved@a{\expandafter\@gobble\string #1}%
     \@ifundefined\reserved@a
         {\edef\reserved@b{\expandafter\@carcube \reserved@a xxx\@nil}%
          \ifx \reserved@b\@qend \@notdefinable\else
            \ifx \reserved@a\@qrelax \@notdefinable\else
              #2%
            \fi
          \fi}%
         \@notdefinable}
\let\@@ifdefinable\@ifdefinable
\long\def\@rc@ifdefinable#1#2{%
  \let\@ifdefinable\@@ifdefinable
  #2}
\def\newenvironment{\@star@or@long\new@environment}
\def\new@environment#1{%
  \@testopt{\@newenva#1}0}
\def\@newenva#1[#2]{%
   \@ifnextchar [{\@newenvb#1[#2]}{\@newenv{#1}{[#2]}}}
\def\@newenvb#1[#2][#3]{\@newenv{#1}{[#2][#3]}}
\def\renewenvironment{\@star@or@long\renew@environment}
\def\renew@environment#1{%
  \@ifundefined{#1}%
     {\@latex@error{Environment #1 undefined}\@ehc
     }{}%
  \expandafter\let\csname#1\endcsname\relax
  \expandafter\let\csname end#1\endcsname\relax
  \new@environment{#1}}
\long\def\@newenv#1#2#3#4{%
  \@ifundefined{#1}%
    {\expandafter\let\csname#1\expandafter\endcsname
                         \csname end#1\endcsname}%
    \relax
  \expandafter\new@command
     \csname #1\endcsname#2{#3}%
     \l@ngrel@x\expandafter\def\csname end#1\endcsname{#4}}

\def\providecommand{\@star@or@long\provide@command}
\def\provide@command#1{%
  {\escapechar\m@ne\xdef\@gtempa{{\string#1}}}%
  \expandafter\@ifundefined\@gtempa
    {\def\reserved@a{\new@command#1}}%
    {\def\reserved@a{\renew@command\reserved@a}}%
   \reserved@a}%

\def\@ifundefined#1{%
  \expandafter\ifx\csname#1\endcsname\relax
    \expandafter\@firstoftwo
  \else
    \expandafter\@secondoftwo
  \fi}

\chardef\@xxxii=32
\mathchardef\@Mi=10001
\mathchardef\@Mii=10002
\mathchardef\@Miii=10003
\mathchardef\@Miv=10004

\newcount\@tempcnta
\newcount\@tempcntb
\newif\if@tempswa\@tempswatrue
\newdimen\@tempdima
\newdimen\@tempdimb
\newdimen\@tempdimc
\newbox\@tempboxa
\newskip\@tempskipa
\newskip\@tempskipb
\newtoks\@temptokena

\long\def\@whilenum#1\do #2{\ifnum #1\relax #2\relax\@iwhilenum{#1\relax
     #2\relax}\fi}
\long\def\@iwhilenum#1{\ifnum #1\expandafter\@iwhilenum
         \else\expandafter\@gobble\fi{#1}}
\long\def\@whiledim#1\do #2{\ifdim #1\relax#2\@iwhiledim{#1\relax#2}\fi}
\long\def\@iwhiledim#1{\ifdim #1\expandafter\@iwhiledim
        \else\expandafter\@gobble\fi{#1}}
\long\def\@whilesw#1\fi#2{#1#2\@iwhilesw{#1#2}\fi\fi}
\long\def\@iwhilesw#1\fi{#1\expandafter\@iwhilesw
         \else\@gobbletwo\fi{#1}\fi}
\def\@nnil{\@nil}
\def\@empty{}
\def\@fornoop#1\@@#2#3{}
\long\def\@for#1:=#2\do#3{%
  \expandafter\def\expandafter\@fortmp\expandafter{#2}%
  \ifx\@fortmp\@empty \else
    \expandafter\@forloop#2,\@nil,\@nil\@@#1{#3}\fi}
\long\def\@forloop#1,#2,#3\@@#4#5{\def#4{#1}\ifx #4\@nnil \else
       #5\def#4{#2}\ifx #4\@nnil \else#5\@iforloop #3\@@#4{#5}\fi\fi}
\long\def\@iforloop#1,#2\@@#3#4{\def#3{#1}\ifx #3\@nnil
       \expandafter\@fornoop \else
      #4\relax\expandafter\@iforloop\fi#2\@@#3{#4}}
\def\@tfor#1:={\@tf@r#1 }
\long\def\@tf@r#1#2\do#3{\def\@fortmp{#2}\ifx\@fortmp\space\else
    \@tforloop#2\@nil\@nil\@@#1{#3}\fi}
\long\def\@tforloop#1#2\@@#3#4{\def#3{#1}\ifx #3\@nnil
       \expandafter\@fornoop \else
      #4\relax\expandafter\@tforloop\fi#2\@@#3{#4}}
\long\def\@break@tfor#1\@@#2#3{\fi\fi}
\def\@removeelement#1#2#3{%
  \def\reserved@a##1,#1,##2\reserved@a{##1,##2\reserved@b}%
  \def\reserved@b##1,\reserved@b##2\reserved@b{%
    \ifx,##1\@empty\else##1\fi}%
  \edef#3{%
    \expandafter\reserved@b\reserved@a,#2,\reserved@b,#1,\reserved@a}}

\let\ExecuteOptions\@gobble

\def\@latex@error#1#2{%
  \errhelp{#2}\errmessage{#1}}

\bgroup\uccode`\!`\%\uppercase{\egroup
\def\@percentchar{!}}

\ifx\@@input\@undefined
 \let\@@input\input
\fi

\def\input{\@ifnextchar\bgroup\@iinput\@@input}
\def\@iinput#1{\input{#1} }

\ifx\filename@parse\@undefined
  \def\reserved@a{./}\ifx\@currdir\reserved@a
    \wlog{^^JDefining UNIX/DOS style filename parser.^^J}
    \def\filename@parse#1{%
      \let\filename@area\@empty
      \expandafter\filename@path#1/\\}
    \def\filename@path#1/#2\\{%
      \ifx\\#2\\%
         \def\reserved@a{\filename@simple#1.\\}%
      \else
         \edef\filename@area{\filename@area#1/}%
         \def\reserved@a{\filename@path#2\\}%
      \fi
      \reserved@a}
  \else\def\reserved@a{[]}\ifx\@currdir\reserved@a
    \wlog{^^JDefining VMS style filename parser.^^J}
    \def\filename@parse#1{%
      \let\filename@area\@empty
      \expandafter\filename@path#1]\\}
    \def\filename@path#1]#2\\{%
      \ifx\\#2\\%
         \def\reserved@a{\filename@simple#1.\\}%
      \else
         \edef\filename@area{\filename@area#1]}%
         \def\reserved@a{\filename@path#2\\}%
      \fi
      \reserved@a}
  \else\def\reserved@a{:}\ifx\@currdir\reserved@a
    \wlog{^^JDefining Mac style filename parser.^^J}
    \def\filename@parse#1{%
      \let\filename@area\@empty
      \expandafter\filename@path#1:\\}
    \def\filename@path#1:#2\\{%
      \ifx\\#2\\%
         \def\reserved@a{\filename@simple#1.\\}%
      \else
         \edef\filename@area{\filename@area#1:}%
         \def\reserved@a{\filename@path#2\\}%
      \fi
      \reserved@a}
  \else
    \wlog{^^JDefining generic filename parser.^^J}
    \def\filename@parse#1{%
      \let\filename@area\@empty
      \expandafter\filename@simple#1.\\}
  \fi\fi\fi
  \def\filename@simple#1.#2\\{%
    \ifx\\#2\\%
       \let\filename@ext\relax
    \else
       \edef\filename@ext{\filename@dot#2\\}%
    \fi
    \edef\filename@base{#1}}
  \def\filename@dot#1.\\{#1}
\else
  \wlog{^^J^^J%
    \noexpand\filename@parse was defined in texsys.cfg:^^J%
    \expandafter\strip@prefix\meaning\filename@parse.^^J%
    }
\fi

\long\def \IfFileExists#1#2#3{%
  \openin\@inputcheck#1 %
  \ifeof\@inputcheck
    \ifx\input@path\@undefined
      \def\reserved@a{#3}%
    \else
      \def\reserved@a{\@iffileonpath{#1}{#2}{#3}}%
    \fi
  \else
    \closein\@inputcheck
    \edef\@filef@und{#1 }%
    \def\reserved@a{#2}%
  \fi
  \reserved@a}
\long\def\@iffileonpath#1{%
  \let\reserved@a\@secondoftwo
  \expandafter\@tfor\expandafter\reserved@b\expandafter
             :\expandafter=\input@path\do{%
    \openin\@inputcheck\reserved@b#1 %
    \ifeof\@inputcheck\else
      \edef\@filef@und{\reserved@b#1 }%
      \let\reserved@a\@firstoftwo%
      \closein\@inputcheck
      \@break@tfor
    \fi}%
  \reserved@a}
\long\def \InputIfFileExists#1#2{%
  \IfFileExists{#1}%
    {#2\@addtofilelist{#1}\@@input \@filef@und}}

\chardef\@inputcheck0

\let\@addtofilelist \@gobble

\def\@defaultunits{\afterassignment\remove@to@nnil}
\def\remove@to@nnil#1\@nnil{}

\newdimen\leftmarginv
\newdimen\leftmarginvi

\newdimen\@ovxx
\newdimen\@ovyy
\newdimen\@ovdx
\newdimen\@ovdy
\newdimen\@ovro
\newdimen\@ovri
\newdimen\@xdim
\newdimen\@ydim
\newdimen\@linelen
\newdimen\@dashdim

\long\def\mbox#1{\leavevmode\hbox{#1}}

\let\@onlypreamble\@gobble

\let\protect\relax

\newdimen\fboxsep
\newdimen\fboxrule

\fboxsep = 3pt
\fboxrule = .4pt

\def\@height{height} \def\@depth{depth} \def\@width{width}
\def\@minus{minus}
\def\@plus{plus}
\def\hb@xt@{\hbox to}

\long\def\@begin@tempboxa#1#2{%
   \begingroup
     \setbox\@tempboxa#1{\color@begingroup#2\color@endgroup}%
     \def\width{\wd\@tempboxa}%
     \def\height{\ht\@tempboxa}%
     \def\depth{\dp\@tempboxa}%
     \let\totalheight\@ovri
     \totalheight\height
     \advance\totalheight\depth}
\let\@end@tempboxa\endgroup

\let\set@color\relax
\let\color@begingroup\relax
\let\color@endgroup\relax
\let\color@setgroup\relax

\let\color@hbox\relax
\let\color@vbox\relax
\let\color@endbox\relax


\begingroup
  \catcode`P=12
  \catcode`T=12
  \lowercase{
    \def\x{\def\rem@pt##1.##2PT{##1\ifnum##2>\z@.##2\fi}}}
  \expandafter\endgroup\x
\def\strip@pt{\expandafter\rem@pt\the}


\def\@input#1{%
  \IfFileExists{#1}{\@@input\@filef@und}{\message{No file #1.}}}

\def\@warning{\immediate\write16}


\def\Gin@driver{dvips.def}
\input graphicx.sty
\resetatcatcode


\resetatcatcode                 

\def\ideq{\equiv}		

\def\sqr#1#2{\vcenter{
  \hrule height.#2pt 
  \hbox{\vrule width.#2pt height#1pt 
        \kern#1pt 
        \vrule width.#2pt}
  \hrule height.#2pt}}


\def\dal{\mathop{\,\sqr{7}{5}\,}}
\def\block{\dal}

\font\openface=msbm10 at10pt
\def\Minkowski     {{\hbox{\openface M}}}
\def\Reals         {{\hbox{\openface R}}}

\def\gto{\mathop
        {\hbox{${\lower3.8pt\hbox{$>$}}\atop{\raise0.2pt\hbox{$\sim$}}$}}}


\def\Caption#1{
\vskip 1cm
\vbox{
 \leftskip=1.5truecm\rightskip=1.5truecm     
 \singlespace                                
 \noindent #1
 \vskip .25in\leftskip=0truecm\rightskip=0truecm 
} 
 \vskip 0.01cm
 \sesquispace}




\phantom{}





\sesquispace
\centerline{\titlefont When is an area law not an area law?}

\bigskip


\singlespace			        

\author{{Anushya Chandran$^{1}$}, {Chris Laumann$^{1,2}$}, {Rafael D. Sorkin$^{1,3}$}}
\address
 {$^{1}$Perimeter Institute, 31 Caroline Street North, Waterloo ON, N2L 2Y5 Canada}
\address
 {$^{2}$Department of Physics, University of Washington, Seattle, WA 98195, USA}
\address
 {$^{3}$Department of Physics, Syracuse University, Syracuse, NY 13244-1130, U.S.A.}
\email{rsorkin@perimeterinstitute.ca}

\AbstractBegins                              
Entanglement entropy is typically proportional to area, 
but sometimes
it acquires an additional logarithmic pre-factor.  
We offer some intuitive explanations for these facts.
\AbstractEnds                                



\sesquispace
\vskip -10pt

\section{} 
Born of the attempt to understand black hole thermodynamics, 
the concept of entanglement entropy 
has since gone out into the wider world and made itself at home 
in milieus ranging from 
conformal field theory to
lattice-fermions 
to causal sets.  
In the course of this odyssey, the so
called {\it\/area law\/} that
entropy is proportional to horizon area has largely maintained its
validity, but not always.

Given a spatial region $B$ 
and its spatial complement ${B}^{\,\prime}$,
the entanglement entropy $S$ between $B$ and $B'$
is said to satisfy an area-law if it
is proportional to the area $A$ of the surface
$\partial{B}$ that separates $B$ from ${B}^{\,\prime}$.
Here, of course, one is using the word ``area'' loosely to denote a
quantity which carries dimensions of $[length]^{d-1}$ when
the spatial dimension is $d$ (the spacetime dimension is $d+1$).
One is also assuming that 
the quantum field (or spin-system, etc.) 
under consideration
is in, or sufficiently near to, its ground state or ``vacuum''.

For $d=1$,
when $B$ is a line-segment or half-line, 
its boundary is reduced to just one or two isolated points,
and a strictly valid area law 
in this case 
would require $S$ to be independent of the size of $B$.
However, this is not always true. 
Rather the putative $[length]^0$ scaling of $S$
for $d=1$
is sometimes
replaced
by a logarithm, as happens for example if we are dealing with a massless field.
Although strictly speaking, 
a logarithmic scaling
is not an area law,
it is completely consistent with the rule of thumb that a quantity which
scales like $x^n$ for generic $n$, often scales like $\log{x}$ when
$n=0$.  In that sense, the logarithm for $d=1$ fits nicely into a
general pattern.

What doesn't fit so nicely, though, is the fact that even in dimensions
greater than $1+1$, the entanglement entropy can acquire an additional
prefactor of the form $\log L$ where $L$ is some characteristic size of
the region $B$.  
This happens in particular when the system in question is 
a collection of free fermions 
in their ground state at nonzero density
(i.e. when they are at zero temperature but nonzero chemical potential).   
Of course a logarithm represents a relatively mild deviation from a
constant, but the discrepancy is big enough that it asks to be explained.
Let us, then,  try to develop an intuition for
when an area law ought to hold, and why.\footnote{$^\star$}
{Although some of the arguments we will give herein may be new to a
 certain extent, this paper is intended primarily as a didactic
 introduction to certain heuristic viewpoints which have arisen over the
 years in relation to entanglement entropy.  In discussing entanglement
 entropy with various people, we have found that depending on which
 community of physicists they belonged to, they have typically been
 familiar with some of these viewpoints, but unfamiliar with others.  By
 bringing them together in one place, we hope to make a more unified
 view available to the community as a whole.  [1]}

Think for example of a spatial region $B$ in the shape of a ball, a disk
or a line-segment, depending on the dimension, and consider a 
massless (gapless) scalar field 
which we take to be in its ``vacuum'' or ``ground state''
and 
whose entanglement 
with ${B}^{\,\prime}$
we wish to evaluate.
One might imagine that
if the field can be decomposed into
quasi-localized ``modes'' 
(``wavelets'', for example [2]),
then the entanglement entropy $S$ will come
primarily from the modes that ``{\it\/straddle\/}'' the boundary $\partial{B}$,
with each such mode contributing
on the order of
one bit of entropy.  
The total entropy would then be proportional to the number of straddling modes.
How many are there?

\vskip 35pt
\vbox{\bigskip
  \centerline {\includegraphics[scale=0.5]{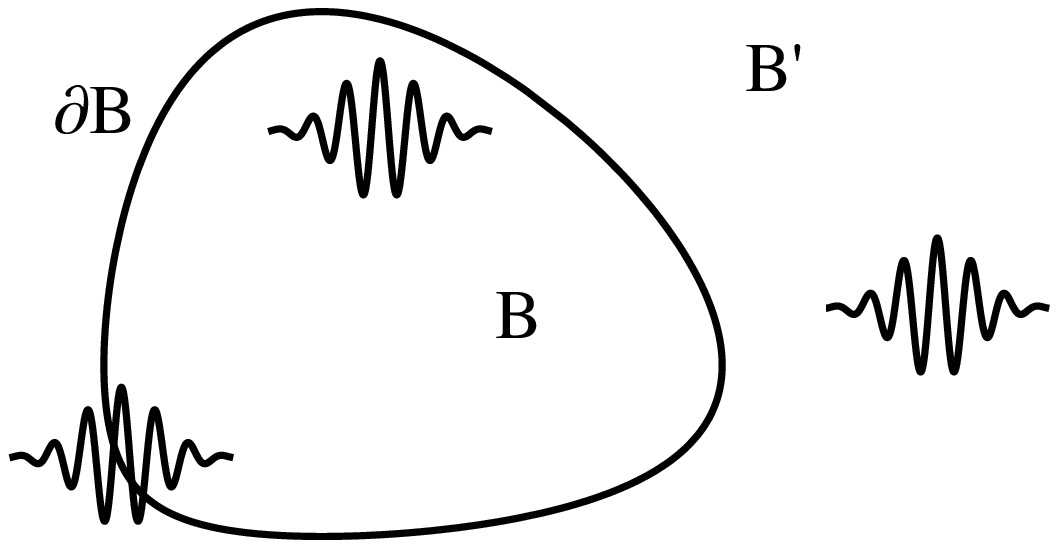}}
  \Caption{{\it Figure 1.} A spatial region $B$ and its boundary $\partial{B}$,
      showing ``modes'' which do and do not straddle $\partial{B}$. }}

Without some restriction or ``cutoff'' there are of course infinitely
many straddling modes of shorter and shorter wavelength $\lambda$.
But if we
limit ourselves to $\lambda>\ell$ for some
$\ell>0$, 
the
number of modes will be finite, 
and we can estimate it as follows, 
assuming for the moment that $d=3$ 
and writing $L$ for the diameter of the ball.
Let $k=2\pi/\lambda$ as usual.  Then in the range $d^3k$
about $k$, the number of modes will be approximately $V\,d^3k/(2\pi)^3$,
where $V$ is the volume of the ball $B$.  Those which straddle
the sphere $\partial{B}$ 
will comprise a fraction $V'/V$ of the total, 
where $V'\approx\lambda A$ 
with $A$ being the area of $\partial{B}$, 
whence the number of straddling modes will be
$\lambda\,A\,d^3k/(2\pi)^3$.  
Integrating this from $k_{min}={2\pi}/L$
up to 
$k_{max}={2\pi}/\ell$,
and assuming that $\ell{\ll}L$,
yields then the area law,
$$
     S \propto N \approx \int\limits_{k_{min}}^{k_{max}} \lambda A {d^3k\over (2\pi)^3}
          = \int\limits^{2\pi/\ell}_{2\pi/L} {2\pi\over k} A {4\pi k^2 dk \over (2\pi)^3}
          = {A \over 2\pi} \left( \left( {2\pi\over \ell} \right)^2 - \left( {2\pi\over L} \right)^2 \right)
          \approx { 2\pi A \over \ell^2}
$$
In spatial dimension $d$ the integrand 
will contain $k^{d-2}\,dk$ 
in place of
$k\,dk\,$,
and the integration will yield instead of a coefficient of 
$(1/\ell^{2}-1/L^{2})$ 
a coefficient of 
$(1/\ell^{d-1}-1/L^{d-1})$, 
except when $d=1$, 
in which case 
the integration
will yield
a coefficient of 
$\log(L/\ell)$.
Thus we expect an area law for $d>1$ and a ``log-corrected area law'' for
$d=1$. 


In dropping the term $1/L^2$, we have
of course 
assumed that $\ell{\ll}L$, 
and under this assumption the above derivation shows 
that almost all of the contributing modes lie very close to the
boundary.  When this is true,
and provided that the boundary is sufficiently smooth,
its detailed shape 
becomes irrelevant, 
and our estimate for the sphere should be valid in general. 

So far we have been 
thinking of
a massless field.
With a nonzero mass $m$,
one would expect entanglement to extend at most over a correlation
length or ``Compton wavelength'', $1/mc$, meaning that only modes with
wavelengths shorter than $1/mc$ should be counted
in estimating the entropy.
Consequently,
instead of a lower limit $k_{min}\sim1/L$, 
our mode-counting integral 
should now have a lower limit of
$\sim\,mc$, 
or just $m$ 
if we set the speed of light $c$ to 1, along with $\hbar$.  
For $d>1$ this change 
to $k_{min}$
makes very little difference, 
provided that $\ell\ll1/m$, 
but for $d=1$, 
it has the effect of converting $\log(L/\ell)$ into 
$\log(1/m\ell)$, a form we will need below.

The analysis 
we have just gone through 
explains why an area law arises in
general, 
and also why 
a logarithmic dependence on $L$
shows up 
in 1+1-dimensions 
when $m=0$.   
But it also
leads us to ask why similar reasoning fails in the
Fermi-surface case.  Before suggesting an answer, 
let us consider 
another 
argument for an area law, 
which in 
one way
is more precise,
but
which applies only when the region in
question is a halfspace and the quantum field in question is
relativistic. 
In that case, one can view the problem from the perspective of a
``Rindler observer'', for whom, 
as is well known, 
the Minkowski vacuum
appears as a thermal state with a local temperature $T$ proportional to the
local redshift and given by $\beta\ideq1/T=2\pi z$ where $z$ is the distance to the
boundary plane (or ``horizon'').  Now a massless field 
in a thermal state of {\it\/uniform\/} temperature $T$ carries an
entropy density proportional to $T^3$ when $d=3$, or to $T^d$ in general.
In our case, the temperature is not uniform, but assuming a local
analysis is approximately valid, we can simply replace $T$ by $1/2\pi z$ and
estimate the total entropy by integrating 
$T^3$
over the region of interest.

Since translation parallel to the boundary is a symmetry 
of the halfspace,
the entropy per unit
boundary area is the relevant quantity, and it will be given by an
integral\footnote{$^\dagger$}
{We henceforth abandon any effort to keep track of numerical prefactors,
 which in any case depend on the details of the cutoff, unless $d=1$.}
%
$\int T^3 dz \sim \int dz/z^3$.
In this situation it seems natural to introduce the UV cutoff by
restricting the range of integration to $z>\ell$, i.e. by ignoring
points closer than $\ell$ to the boundary.  If we do so then there will
correspond to a boundary-portion of area $A$ an entropy of 
$A \int_{1/\ell}^{1/L} dz/z^3 \sim A/\ell^2$, as before.  For general $d>1$
the same reasoning yields $A/\ell^{d-1}$, 
while for $d=1$ it yields
$\log(L/\ell)$.  Thus we reach exactly the same conclusions as before.
In this case, however, it is obvious that the analysis is inapplicable
when a Fermi surface is involved, because the nonzero density of
electrons (or other fermions) breaks the symmetry of Lorentz-boosts (at
least naively).

To progress further it seems necessary to distinguish more carefully
between the component of the momentum-vector which is parallel to
the boundary surface (and which we will call {\it tangential\/})
and the component which is orthogonal to it
(and which we will call {\it\/longitudinal\/}).
In order to have a definite setup to work
with, 
let us consider 
a free relativistic scalar
field of mass $m\ge0$ in a halfspace.  Also, 
in order to 
distinguish between the 
tangential and longitudinal
projections of the wave-vector without
introducing 
cumbersome subscripts
like $k_{\parallel}$ and $k_{\perp}$,
let us 
use the letter 
``$p$'' for the tangential wave-vector, 
while reserving ``$k$'' for
the longitudinal component.  (See Figure 2.)


\vbox{
   \bigskip

  \centerline {\includegraphics[scale=0.5]{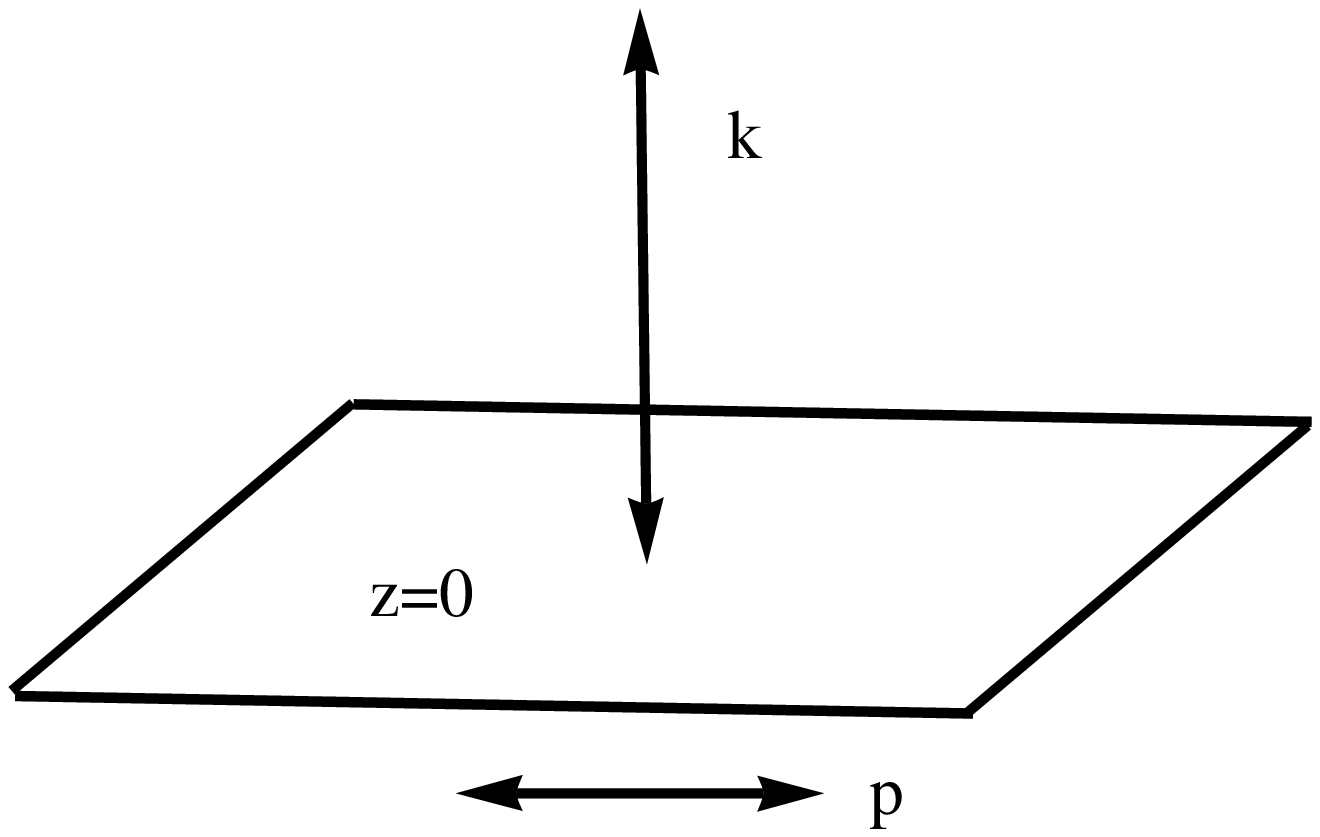}}
  \Caption{{\it Figure 2.} Flat boundary surface and the ``tangential'' and ``longitudinal'' wave-vectors $p$ and $k$. }}

In view of the translational symmetry along the boundary, it is
evident that 
modes with different values of $p$ will not interact with each other 
and will contribute independently to the entanglement.  
Thus it suffices to estimate the contribution from each $p$ and 
add them up to obtain the total entropy $S$ per unit boundary area.
In this way, 
the problem is reduced to series of $1+1$-dimensional problems
parameterized by the tangential momentum-vector.

A mode with tangential momentum $p$ 
takes the form $\phi(t,x,z)=e^{ip{\cdot}x} f(t,z)$ 
where $z$ is the longitudinal coordinate as before, 
and $x$ coordinatizes the boundary surface, $z=0$.
Substituted in the wave equation, $(\block-m^2)\phi=0$, 
this Ansatz produces the 2D equation,
$$
    -{\partial^2 f \over \partial t^2} + {\partial^2 f \over \partial z^2} - (m^2+p^2) f = 0 \ .
      \eqno(1)
$$
In other words, it produces the  $d=1$ wave equation with $m^2+p^2$
playing the role of $m^2$.  The important thing to notice here is that,
{\it with the sole exception of $\;p=0$, we are always in the massive (gapped) case}, 
even when the field overall is massless.

Now a massive free field in  $d=1$ is 
one of the examples we've already encountered.  
Based on our earlier analysis of that case,
we expect (and in fact we know) that 
the entropy will scale as $\log(1/m\ell)$.\footnote{$^\flat$}
{Even without any real thought we can be confident on dimensional
 grounds that $S$ in this case will be some function, $f(m\ell)$.  All
 that the more detailed analysis contributes is the particular form of
 $f$.  That this form is logarithmic guarantees via (2) that the
 entropy per unit area is finite, but that would have been true even if
 $f(x)$ had diverged as a power law $1/x^\gamma$ with $\gamma < d-1$.}
Given this, it is an easy matter to
deduce the scaling of the total
entropy (and therewith the area law).  
Returning for simplicity to $m=0$,
we obtain from tangential momentum $p$ an effective mass of 
$\sqrt{m^2+p^2}=\sqrt{p^2}=|p|$ and a corresponding entropy of
$S(p)\sim\log(1/\ell|p|)$, 
valid for $p\ll1/\ell$. 
(For $p\gto1/\ell$, $S(p)$ falls rapidly to zero.) 
Integrating over $p$ then yields (in $d=3$) an
entropy per unit area of
$$
   \int\limits^{} {d^2p\over (2\pi)^2} \,  S(p) \approx \int\limits_0^{1/\ell} p\, dp\, \log(1/\ell p) = {1 \over 4 \ell^2} \ ,
   \eqno(2)
$$
or corresponding to a portion of the boundary of area $A$,
an entropy $\sim\, A/\ell^2$. 
Thus, we confirm our earlier conclusion in a more rigorous manner, 
and more importantly, 
in a manner which lends itself 
to being carried over to the case of a Fermi surface.  

In light of the foregoing analysis,
it seems possible to put one's mathematical finger on the reason why  
a logarithmic divergence 
(and hence a departure from a simple area law)
shows up in $d=1$, but not in higher dimensions.  
If we take equation (2) as our starting point, 
then it's evident that 
a logarithm is mathematically present in all dimensions, 
but since it is associated only with vanishing tangential momentum, 
its contribution to the entanglement entropy 
is ``drowned out'' when $d\ge2$.
As a result,
one obtains an area law 
in higher dimensions,
whether or not the field is massless.\footnote{$^\star$}
{The integrand $\log(1/\ell p)$ in (2) is not correct all the way
 down to $p=0$.  Rather it goes over to the form $\log(L/\ell)$ for
 $p\ll 1/L$.  However that portion of the integral yields (approximately) only
 $\int_{0}^{1/L} p\, dp\, \log(L/l) = \log(L/l) / 2L^2$, which is
 negligible in comparison to $1/4\ell^2$ when $\ell\ll L$.}



Although the logarithmic scaling in $d=1$ is well established, and
although we have tried above to provide qualitative explanations for it, one
might still wonder whether there is anything more to be gleaned from it
on an intuitive, or even a quantitative level.
To that end, we would like to digress briefly 
on
the massless scalarfield in $d=1$.
When $m=0$, 
the entropy can depend only on $L/\ell$,
or possibly the ratio 
of either $L$ or $\ell$ 
to some further IR scale that might enter the problem.  
The only real question, then, is why the
dependence turns out to be logarithmic.  Such a dependence fits nicely
into the $\ell^{1-d}$ hierarchy, but can one give a less formal reason?

More specifically, can we relate the logarithmic scaling of the entropy
to the ``infrared nonlocality'' that accompanies masslessness in
1+1-dimensions?
A first symptom of this ``nonlocality'' is the fact that,
rather than falling off with separation, 
the Green function 
for a massless scalar field 
grows logarithmically.
A second symptom, closely related to the first, is that
the Minkowski vacuum fails to exist when $m=0$.
(Neither in full $\Minkowski^2$ nor in $S^1 \times \Reals$ is there a
 normalizable state of minimum energy.\footnote{$^\dagger$}
{For this reason, we believe that most computations claiming to evaluate
 the vacuum entanglement entropy in $1+1$ dimensions are incomplete and
 need to be corrected.  
 Indeed, unpublished numerical simulations [3] strongly
 suggest that in these two cases the entanglement entropy will diverge
 in the absence of some sort of IR regulator 
 (the divergence being logarithmic in the regulator).
 In the analysis above, we have implicitly imposed such a regulator,
 for example Dirichlet conditions at one end of the spatial
 interval $[0,L]$, or as in [4] the choice of
 ``SJ-vacuum'' in some large ``causal diamond'' containing $[0,L]$.}
 Such a state does exist when the spacetime is a halfspace or a strip,
 with Dirichlet boundary conditions.)
These symptoms
must be related to the logarithmic scaling of the entropy, but
can one be more precise about how?
(Is there, perhaps, a more detailed breakdown of how the entanglement
entropy arises
that might also carry over to other cases of interest?
For example, could one identify 
some sort of ``falloff of entanglement with distance'' 
which would 
yield the entropy
when integrated over all
pairs of points on either side of the boundary?  If so, and if the
entanglement fell off as
$1/r^2$, then 
logarithmic scaling would follow.)
(In a rather different vein, reference [5] suggests 
that the logarithmic scaling in $d=2$ might be connected to a
similar looking corner-contribution in $d=3$.)

The analysis that led us to equation (2) parallels 
the original proof of the area law in [6] and [7].  
As in those papers,  
most of our discussion here so far
has been limited to the case
of a free scalar field 
(in its vacuum state or sufficiently near to it).
Let us now try to 
carry our analysis over
a ``gas'' of electrons,
treated as non-interacting,
and
held at zero temperature 
but {\it\/nonzero\/}, nonrelativistic density.

For the one-electron states, we can do 
exactly the same decomposition into
tangential modes as before 
and thereby reduce the calculation again to
a succession of $d=1$ problems.  (Equivalently we are decomposing the
field operator $\psi$ that creates the 1-electron states.) 
Making for
the 1-electron Schr{\"o}dinger equation the same
substitution
as we made earlier for the Klein-Gordon equation,
then teaches us 
that
the effect of the tangential momentum $p$ 
in the present, nonrelativistic case 
is just to shift the zero of energy upward by $p^2/2m$, 
as if we had added a constant potential term, $V=p^2/2m$,
to the Hamiltonian. 

\vskip 1.1cm
\vbox{\bigskip
  \centerline {\includegraphics[scale=0.5]{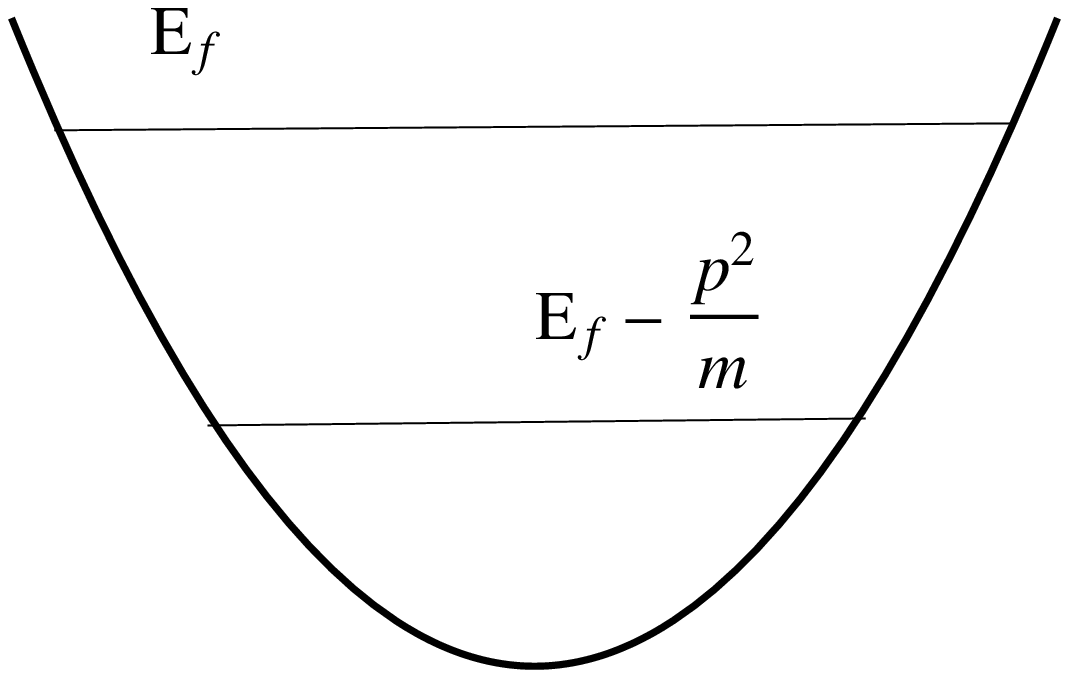}}
  \Caption{{\it Figure 3.}  The effective Fermi level in $k$ is shifted downard by $p^2/2m$}}

Consequently, 
for a given value of $p$,  
the $d=1$ free-particle energy levels will only be filled up to a maximum
energy of
$E_f(p)=E_f-p^2/2m$, 
i.e. they will be filled up to a 
longitudinal
momentum $k$ such that $k^2/2m=E_f(p)$.
The Fermi level $E_f$ is effectively shifted downward, but otherwise
nothing changes.
(Relativistically, the dispersion relation itself would also become a
function of $p$, but it would remain convex.) 
Thus each 1d problem has its own Fermi ``surface'', its location being
set by $p^2$.  
If we can solve for the entropy $S(p)$ in this situation, we have only to
integrate the result, exactly as before.

Now for each given $p$ 
(except very near $E_f(p)=0$)
we can approximate the
dispersion relation 
in the neighborhood of 
$E=E_f(p)$ by a straight line.  Near
$k=\sqrt{2m E_f(p)}$ our 1d problem thus looks like that of a
relativistic massless field, for which we already know the answer.  If
we accept this picture 
[this is perhaps the weakest link in the argument!],
we can immediately conclude that $S(p)$ scales like $\log(L/\ell)$.
But unlike 
in equation (2),
this scaling law is
independent of the tangential momentum $p$, 
and  that being the case, the
integral $\int dp\,p\,S(p)$ will scale the same way.
%
The logarithm is thus not drowned out and remains
as an overall factor when we sum up the contributions, 
while the summation itself produces a factor of $1/\ell^2$ as before. 
The entanglement entropy will therefore
scale like 
$\log(L/\ell) \, A/\ell^2$ 
rather than just $A/\ell^2$.

It remains to determine, at least in order of magnitude, the value of
the cutoff $\ell$.  In a continuum quantum field theory, any cutoff must
come from outside the theory.  Either it is put in by hand, or it
emerges from some deeper reality.  In the earlier case of a scalar
field, our cutoffs were ad hoc and introduced by hand, but in the
present situation the deeper reality is known, and it implies that the
continuum approximation will fail when one reaches a distance-scale
comparable to the mean separation between electrons.  This therefore is
the appropriate value of $\ell$, and since it coincides with the scale
set by the ``Fermi momentum'' $p_F$, we can also think of $\ell$ as
representing $1/p_F$.

With the determination of $\,\ell\,$ 
our discussion of the fermion gas
has reached its destination.
But what might raise doubts about our
explanation of the logarithmic prefactor 
is that in concentrating on the
approximately linear dispersion relation near the effective Fermi
surface, it seemingly attributes the entanglement entropy entirely to
the ``shallow electrons'' in the Dirac sea, whereas it seems clear that
``deeper electrons'' must be contributing most of the entropy since
they are the majority (at least in a 1d context).  What perhaps
rescues the explanation however, is that the deeper electrons do in fact
make their presence felt via their influence on the cutoff,
because
it's the total number-density $n$ 
that determines $\ell$ 
through the relation $\ell\sim n^{-1/3}$. 
In this way the proposed explanation attains 
a certain logical coherence.

Before concluding we would like to comment on an aspect of our treatment
that in a certain sense is unsatisfactory, namely its appeal to space as
opposed to spacetime, with the consequent use of a cutoff that fails to
be Lorentz invariant.  For the purposes of condensed matter, this is not
a problem, of course, but in the context of relativistic quantum field
theory, or of black hole entropy, a more four-dimensional treatment
would be desirable.  Such a method was introduced in [8] and
illustrated in [4] by a $1+1$-dimensional calculation of
entanglement entropy.  Taking this as a point of reference, it would be
interesting to ask whether the intuitive explanations offered above
could also be cast into spatio-temporal form.

\bigskip

We thank Roger Melko and Guifre Vidal for discussing these ideas with
us.  Special thanks go to to Sumati Surya for preparing the diagrams.
This research was supported in part by NSERC through grant RGPIN-418709-2012.
This research was supported in part by Perimeter Institute for Theoretical
Physics.  Research at Perimeter Institute is supported by the Government of
Canada through Industry Canada and by the Province of Ontario through the
Ministry of Research and Innovation.

\ReferencesBegin                             


\ref [1]
For recent reviews of entanglement entropy and the area law see [9] and [10].  
\sepref
On the area law for a free scalar field see [6] and [7],
also [11].  Note, however, that contrary to the false claim in
[11] (``Note added''), the analysis in [7] was
not limited to $m^2>0$.  On the contrary, all values of $m^2$ were
treated together in [7], and the emphasis, if any, was precisely
on the massless case.
\sepref
For a cubic region within a cubic lattice of harmonic oscillators see [12].
\sepref
Logarithmic scaling of entanglement entropy at a quantum critical point
in $1+1$-dimensions is discussed in [13] and [14].
\sepref
For the logarithmic modification of the area law in the presence of a
Fermi surface see [15], [16], and [17].
\sepref
For a logarithmic prefactor coming from a ``Bose surface'' rather than ``Fermi surface''
see [18].
\sepref An initial computation of entanglement entropy in a causal set
is described in [19]

\ref [2]  A computation of entanglement entropy
  using wavelets can be found in [20].  For a general
  introduction to wavelets, see [21]

\ref [3] Yasaman Yazdi (unpublished)

\ref [4] Mehdi Saravani, Rafael D.~Sorkin, and Yasaman K.~Yazdi ``Spacetime Entanglement Entropy in 1+1 Dimensions''
  \arxiv{1205.2953}
  \journaldata{Class. Quantum Grav.}{31}{214006}{2014} 
   available at http://stacks.iop.org/0264-9381/31/214006

\ref [5] Pablo Bueno, Robert C. Myers, and William Witczak-Krempa, ``Universal corner entanglement from twist operators''
     http://arxiv.org/pdf/1507.06997.pdf

\ref [6] Rafael D. Sorkin, ``On the Entropy of the Vacuum Outside a Horizon'',
  in B. Bertotti, F. de Felice and A. Pascolini (eds.),
  {\it Tenth International Conference on General Relativity and Gravitation (held Padova, 4-9 July, 1983), Contributed Papers}, 
  vol. II, pp. 734-736
  (Roma, Consiglio Nazionale Delle Ricerche, 1983), \lbr
  \eprint{http://www.pitp.ca/personal/rsorkin/some.papers/31.padova.entropy.pdf}

\ref [7] Luca Bombelli, Rabinder K.~Koul, Joohan Lee and Rafael D.~Sorkin, ``A Quantum Source of Entropy for Black Holes'', 
  \journaldata{Phys. Rev.~D}{34}{373-383}{1986},
   reprinted in: B.-L.~Hu and L.~Parker (eds.) {\it Quantum Theory in Curved Spacetime} (World Scientific, Singapore, 1992)

\ref [8] Rafael D.~Sorkin, ``Expressing entropy globally in terms of (4D) field-correlations'',
 \journaldata{J. Phys. Conf. Ser.}{ 484}{ 012004} {2014}
 (Proceedings of the Seventh International Conference on Gravitation and Cosmology [ICGC], held December 2011 in Goa, India)
 \arxiv{1205.2953}  
 \lbr
 \eprint{http://www.pitp.ca/personal/rsorkin/some.papers/143.s.from.w.pdf} 

\ref [9]
  Luigi Amico, Rosario Fazio, Andreas Osterloh, and Vlatko Vedral,
  ``Entanglement in Many-Body Systems''  
  \journaldata{Rev. Mod. Phys.} {80}{517} {2008}

\ref [10]
  Eisert, Cramer, and Plenio,
  ``Area laws for the entanglement entropy - a review'' 
  \journaldata {Rev. Mod. Phys.}{82}{277}{2010}

\ref [11]
Mark Srednicki, ``Entropy and Area'',
 \journaldata{Phys.~Rev.~Lett.}{71}{666-669}{1993}
 \eprint{hep-th/9303048}

\ref [12]
M.B. Plenio, J. Eisert, J. Dreissig, and M. Cramer,
``Entropy, entanglement, and area: analytical results for harmonic lattice systems''
\journaldata{Phys. Rev. Lett.}{94}{060503}{2005}

\ref [13]
G. Vidal, J. I. Latorre, E. Rico, A. Kitaev,
``Entanglement in quantum critical phenomena'', 
\journaldata{Phys. Rev. Lett.} {90} {227902} {2003}

\ref [14]
Pasquale Calabrese and John Cardy,
``Entanglement entropy and quantum field theory'', 
\journaldata{J. Stat. Mech: Theory and Experiment}{}{P06002}{2004}

\ref [15] Dimitri Gioev and Israel Klich, ``Entanglement Entropy of Fermions in Any Dimension and the Widom Conjecture''
\journaldata{Phys. Rev. Lett.} {96} {100503}{2006}

\ref [16] Michael M. Wolf, ``Violation of the Entropic Area Law for Fermions'',
\journaldata{Phys. Rev. Lett.} {96} {010404}{2006}

\ref [17] B. Swingle, ``Entanglement Entropy and the Fermi Surface'', 
  \journaldata{Phys. Rev. Lett.}{105}{050502}{2010}

\ref [18] Hsin-Hua Lai, Kun Yang, and N. E. Bonesteel,
  ``Violation of the Entanglement Area Law in Bosonic Systems with Bose Surfaces: Possible Application to Bose Metals''
  \journaldata{Phys. Rev. Lett.}{111}{210402}{2013}

\ref [19] Yasaman Yazdi, et al. (in preparation)

\ref [20] Gavin K. Brennen, Peter Rohde, Barry C. Sanders, Sukhwinder Singh, 
  ``Multi-scale quantum simulation of quantum field theory using wavelets'', \lbr
   http://journals.aps.org/pra/pdf/10.1103/PhysRevA.92.032315,
   arXiv:1412.0750

\ref [21]
     https://en.wikipedia.org/wiki/Multiresolution\_analysis; \lbr
     and
     https://en.wikipedia.org/wiki/Wavelet

\end


(prog1 'now-outlining
  (Outline* 
     "\f"                   1
      "
      "
      "
      "
      "\\Abstract"          1
      "\\section"           1
      "\\subsection"        2
      "\\appendix"          1       ; still needed?
      "\\ReferencesBegin"   1
      "
      "\\ref "              2
      "\\end